# Automatic Mouse Embryo Brain Ventricle & Body Segmentation and Mutant Classification From Ultrasound Data Using Deep Learning

Ziming Qiu[1], Nitin Nair[1], Jack Langerman[2], Orlando Aristizábal[3,4], Jonathan Mamou[3], Daniel H. Turnbull[4], Jeffrey A. Ketterling[3], Yao Wang[1]

1. Electrical and Computer Engineering, Tandon School of Engineering, New York University, Brooklyn, USA
2. Computer Science, Tandon School of Engineering, New York University, Brooklyn, USA
3. F. L. Lizzi Center for Biomedical Engineering, Riverside Research, New York, USA
4. Skirball Institute of Biomolecular Medicine, New York University School of Medicine, New York, USA

{zq415, nn1174, jackmlangerman, yw523}@nyu.edu, {orlando.aristizabal, daniel.turnbull}@med.nyu.edu {JMamou, Jketterling}@riversideresearch.org

*Abstract*— High-frequency ultrasound (HFU) is well suited for imaging embryonic mice *in vivo* because it is non-invasive and real-time. Manual segmentation of the brain ventricles (BVs) and whole body from 3D HFU images is time-consuming and requires specialized training. This paper presents a deep-learning-based segmentation pipeline which automates several time-consuming, repetitive tasks currently performed to study genetic mutations in developing mouse embryos. Namely, the pipeline accurately segments the BV and body regions in 3D HFU images of mouse embryos, despite significant challenges due to position and shape variation of the embryos, as well as imaging artifacts. Based on the BV segmentation, a 3D convolutional neural network (CNN) is further trained to detect embryos with the *Engrailed-1* (*En1*) mutation. The algorithms achieve 0.896 and 0.925 Dice Similarity Coefficient (DSC) for BV and body segmentation, respectively, and 95.8% accuracy on mutant classification. Through gradient based interrogation and visualization of the trained classifier, it is demonstrated that the model focuses on the morphological structures known to be affected by the *En1* mutation.

## I. Introduction

To investigate how genetic mutations disrupt the body plan during embryonic development, the mouse is commonly used as an animal model because of its high degree of homology with the human genome. In particular, we are interested in the *Engrailed-1* (*En1*) mutation which is lethal *in utero*. One of the key methods for detecting *En1* and similar mutations is to observe how they manifest themselves during embryonic development as variation in the shape of the brain ventricle (BV) and other parts of the body in 3D views [1]. High-frequency ultrasound (HFU) is well suited for imaging embryonic mice because it is real-time, non-invasive and can still provide high

The research described in this paper was supported in part by NIH grant EB022950.

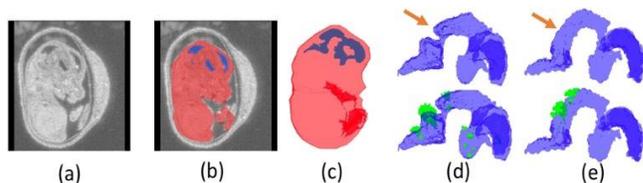

Fig. 1: (a) B-mode slice. (b) BV (blue) and body (red) segmentation. (c) 3D rendering. BVs in (d) normal and (e) mutant embryos. Arrows indicate difference between normal and mutant BVs. Green points indicate regions that maximally influence predictions of the trained mutant classification model.

resolution volumetric datasets [2]. However, manual segmentation of the BV and body (Fig. 1 (b),(c)) from 3D HFU volumes is time-consuming and requires specialized training. Therefore, it is essential to develop fully automatic segmentation and mutant classification algorithms [3].

A graph-based algorithm, Nested Graph Cut (NGC) [4], was first proposed to segment the BV in a HFU mouse embryo head image (manually cropped from the acquired HFU whole-body image). Although a subsequent NGC based framework [5] achieved good performance with 36 HFU whole-body images (Dice Similarity Coefficient (DSC) of 0.892), subsequent testing on a larger data set (111 unseen images) yielded a much lower DSC of 0.7119. The lower performance was likely due to overfitting because the structure and parameters of the framework were tuned to the whole-body images by trial-and-error.

Inspired by the enormous success of the application of deep-learning in computer vision tasks, the biomedical image analysis community has started adopting deep-learning methods for classification, localization and segmentation [6][7]. A deep-learning based framework for BV segmentation was developed in [8] which surpassed NGC based framework in terms of mean DSC and robustness.

Inspired by the success of the work in [8], our framework was extended to segment the body surface using a deep-learning based framework. The challenges for body segmentation are similar to those for BV segmentation, except that the imbalance between the background and the foreground is not as extreme (i.e., the body makes up around 10% of the whole volume on average). Therefore, the localization step is not necessary for body segmentation. A segmentation network (with the same structure as the 3D fully convolution network (FCN) used for BV segmentation in [8]) is trained that can segment a small 3D patch into body and background. The trained network is then applied convolutionally to overlapping 3D patches in the entire volume. Segmentation results in the overlapping regions are determined by taking average of predictions for all overlapping patches.

Next, an automatic mutant classification model is developed. By observing the 3D BV segmentation, it is easy for experts to tell whether a particular mouse embryo is mutant or not (Fig. 1 (d),(e)). This leads to the conclusion that, it is possible to develop a classification model using the BV segmentation map alone.

## II. METHODS

### A. Brain Ventricle Segmentation

Because the BV makes up less than 0.5% of the whole volume, a localization step is used before performing segmentation. In [8], a fully automated framework was proposed consisting of two modules: BV localization and segmentation. More specifically, a 10-layer volumetric VGG-style convolutional neural network (CNN) [9] is used to classify each sliding window (a 3D patch of size $128 \times 128 \times 128$) into two classes: containing the BV or not. At test time, there are generally multiple windows classified as containing the BV. We take the mean position of the center of all the positive sliding windows as the center of the detected BV window. To segment the detected window into BV or background, the 3D FCN illustrated in Fig. 2 was adopted (the network structure design and training are detailed in [8]).

### B. Body Segmentation

For body segmentation this work employs a sliding window based algorithm for body segmentation as illustrated in Fig. 3. Specifically, a $160 \times 160 \times 160$ sliding window with a step size of 16 is used to extract a large number of sub-volumes for training. Inside each extracted sliding window, the same FCN used for BV segmentation (Fig. 2) is used to perform body segmentation. At test time, a weighted average of the predictions for overlapping regions covered by multiple adjacent sliding windows is used. Higher weights are given to those windows where this region is in the center, and lower weights to windows where this region is close to the window boundary. This is because prediction results in the central regions tend to be more accurate, as they are able to exploit larger contexts than those close to the boundaries.

### C. Mutant Classification and Visualization

The fact that human experts can perform mutant classification by just looking at the BV segmentation, implies that it is possible to use the BV segmentation map (the output of [8]) to train a CNN for mutant classification. First, Principle Component Analysis (PCA) is performed to rotate the BV segmentation maps into a canonical pose so as to reduce irrelevant variations in the input to the CNN. Indeed, this pre-processing helps improve the classification performance of the trained network in our experiments. Then, a 9-layer volumetric VGG-style CNN (see Fig. 4) is trained using the rotated BV segmentation. Finally, we interrogate the trained network by propagating gradient of the classification output back to the input BV segmentation image [10]. 20% of the maximum gradient value is used as the threshold to obtain a binary saliency image (Fig. 6). These saliency images serve as the explanations of the classifier's decisions.

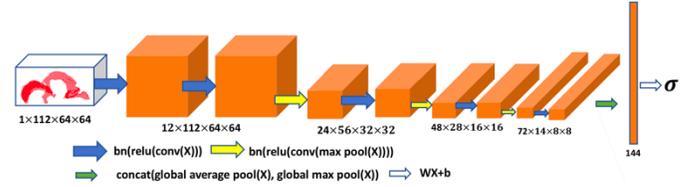

Fig. 4: Pictorial representation of the mutant classification method.

## III. RESULTS AND DISCUSSIONS

All the image data were acquired *in utero* using a 5-element 40-MHz annular array and the voxel size was $50 \times 50 \times 50$ $\mu m$. As illustrated in Table I, a trained expert manually segmented 370 volume images to label the BV voxels; 259 volumes were used for training and 111 for testing in [8]. Among the 370 volumes with manual BV segmentation, a trained expert manually segmented the body for 153 images, among which 107 were used for training and the remaining 46 for testing. Because the BV segmentation algorithm performed satisfactorily, the algorithm was applied on the additional unlabeled images and 196 BV segmentations were hand-picked. This process resulted in a total of 566 images with BV segmentation: 103 mutant and 463 normal (Table I). 6-fold cross validation was used to develop and evaluate the mutant classification algorithm.

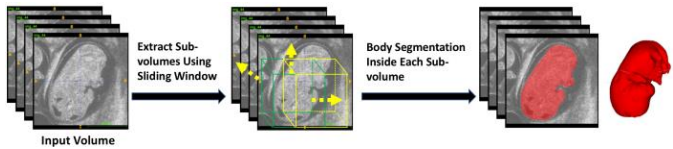

Fig. 3: Pictorial representation of the body segmentation method.

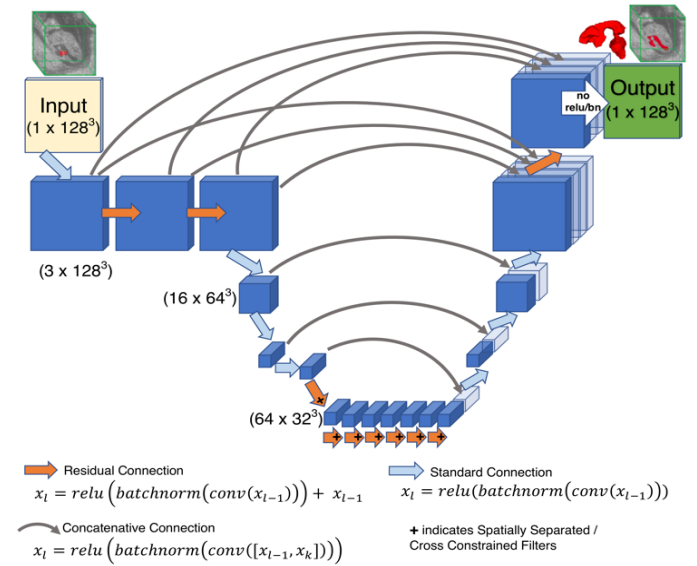

Fig. 2: Pictorial representation of the segmentation network. $x_l$ = the $l_{th}$ layer activations. $x_0$ = input volume.

For segmentation and mutant classification, PyTorch [11] was used to implement the deep neural networks. ITK-SNAP [12] was used to visualize segmentation results.

TABLE I. DATA SET STATICSTICS AND CORRESPONDING TESTING RESULTS

| BV Segmentation | Body Segmentation | Mutant Classification |
|---|---|---|
| 370 manual segmentation, 196 verified automatic segmentation | 153 manual segmentation | 566 images: 103 mutant and 463 normal |
| 259 for training | 107 for training | 6-fold cross validation |
| 111 for testing | 46 for testing | |
| 0.896 DSC | 0.925 DSC | 0.958 accuracy |

### A. BV Segmentation Results

The system was trained on 259 HFU images with manual BV segmentation and achieved a DSC score of 0.896 on the unseen 111 volume test set. More details and sample results can be found in [8].

### B. Body Segmentation Results

For body segmentation, the network was trained on 107 volumes and tested on 46 unseen images achieving 0.925 DSC. As shown in Fig. 5, each image had a different body orientation, shape, image quality and contrast. However, our proposed body segmentation framework can still yield competitive predicted body segmentation when compared to manual segmentation. Due to the low image quality in Fig. 5 (c), the corresponding manual body segmentation did not look like an embryonic mouse body while the predicted body segmentation makes more sense. In Fig. 5 (d), the predicted body segmentation had low DSC because we do not have enough training images with this contrast. We attribute the robustness of our framework to fully convolutional design of the segmentation network which enabled end-to-end and pixel-to-pixel training.

### C. Muant Classification and Visualization Results

Using the BV segmentation algorithm, the BV data set was expanded to 566 images: 103 mutant and 463 normal. Finally, a 3D CNN was trained for mutant classification achieving 95.8% accuracy (more detailed results are shown in Table II). More importantly, the saliency maps of the trained classifier demonstrate that the model focused on the mid-hindbrain region where *En1* mutation is known to cause loss of brain tissue thereby leading to thickening of the BV (Fig. 6).

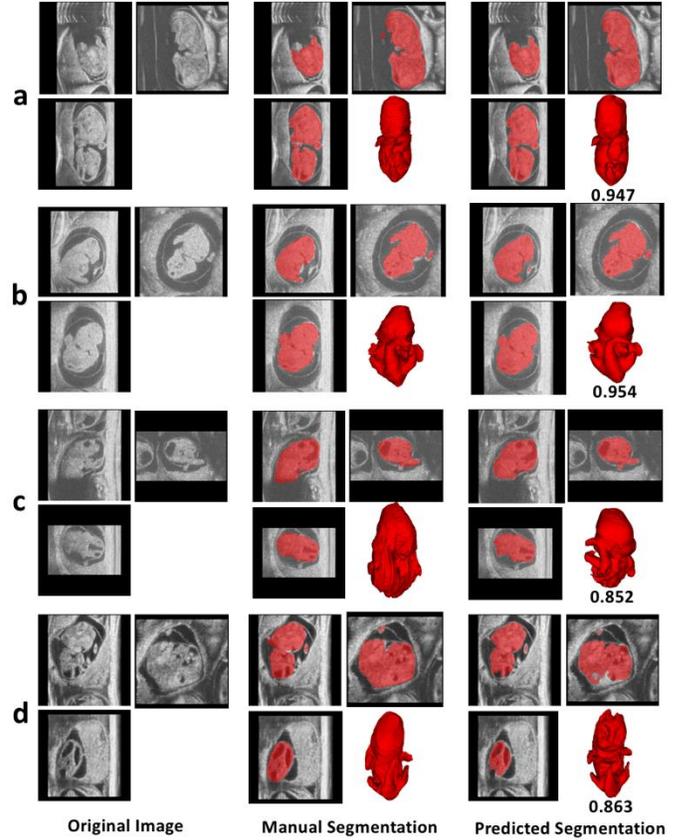

Fig. 5: Visualization of 4 randomly selected body segmentation examples. Each image is shown with coronal (top left), transverse (top right), sagittal (bottom left) and 3D body (bottom right) views. The number below the 3D body view is the corresponding Dice Similarity Coefficient. Note (c) has poor image quality and (d) has different image contrast.

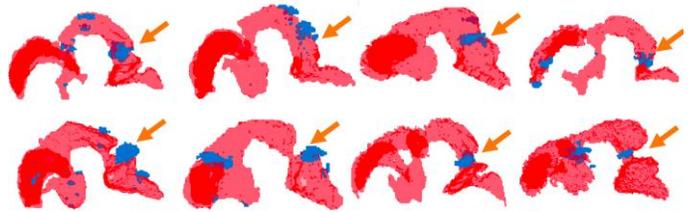

Fig. 6: The visualization of the trained classifier. The first row is mutant BV and the second row is normal BV. Blue points indicate regions that maximally influence predictions of the trained classifier. Orange arrows point to the mid-hindbrain region.

TABLE II. MUTANT CLASSIFICATION RESULTS SUMMED OVER VALIDATION SAMPLES WITH 6-FOLD CROSS VALIDATION. AVERAGE ACCURACY = 0.958.

| Predict / True | Mutant | Normal |
|---|---|---|
| Mutant | 92 | 11 |
| Normal | 12 | 451 |

## IV. CONCLUSION

The proposed fully-automatic, deep-learning based methods for BV and body segmentation are highly promising, achieving high DSC of 0.896 and 0.925 for BV and body segmentations, respectively, on test data. The proposed deep-learning based method for mutant detection from the BV segmentation maps also achieve a high average validation accuracy of 0.958 over 6 cross validation folds. More importantly, the trained classification model is shown to differentiate between mutant and wild-type mouse embryos by focusing on the region where the phenotype associated with the *En1* mutation typically manifests. This suggests that the proposed pipeline can serve as a template, both to aide in the characterization of unknown phenotypes associated with a known gene, and more generally in the use of data driven "black box" predictive models not for prediction, but as tools explain an underlying phenomenon. In conclusion, our segmentation and mutant classification algorithms could be invaluable in streamlining development biology studies.